\documentclass[twocolumn]{aastex701}

\usepackage{float}
\usepackage{comment}
\usepackage[T1]{fontenc}

\newcommand\kms{\rm km\ s^{-1}}

\newcommand\isis{${\rm ^{28}SiS}\ J{=}16{-}15$}
\newcommand\jsis{${\rm ^{30}SiS}\ J{=}18{-}17$}

\begin{document}

\title{ A Protoplanet Candidate in the PDS 66 Disk Indicated by Silicon Sulfide Isotopologues }

\author[0000-0001-8002-8473,sname='Yoshida']{Tomohiro C. Yoshida}
\altaffiliation{These authors contributed equally to this work.}
\affiliation{National Astronomical Observatory of Japan, 2-21-1 Osawa, Mitaka, Tokyo 181-8588, Japan}
\email[show]{tomohiroyoshida.astro@gmail.com}

\author[0000-0002-2692-7862,sname='Alarc\'{o}n']{Felipe Alarc\'{o}n}
\altaffiliation{These authors contributed equally to this work.}
\affiliation{Dipartimento di Fisica, Universit\`{a} degli Studi di Milano, Via Celoria 16, 20133 Milano, Italy}
\email[show]{felipe.alarcon@unimi.it}

\author[0000-0001-7258-770X]{Jaehan Bae}
\email{jbae@ufl.edu}
\affiliation{Department of Astronomy, University of Florida, Gainesville, FL 32611, USA}

\author[0000-0002-7695-7605]{Myriam Benisty}
\email{benisty@mpia.de}
\affiliation{Max-Planck Institute for Astronomy (MPIA), Königstuhl 17, 69117 Heidelberg, Germany}

\author[0000-0003-1958-6673]{Kiyoaki Doi}
\email{doi.kiyoaki.astro@gmail.com}
\affiliation{Max-Planck Institute for Astronomy (MPIA), Königstuhl 17, 69117 Heidelberg, Germany}

\author[0000-0003-4689-2684]{Stefano Facchini}
\email{stefano.facchini@unimi.it}
\affiliation{Dipartimento di Fisica, Universit\`a degli Studi di Milano, Via Celoria 16, 20133 Milano, Italy}

\author[0000-0003-1413-1776]{Charles J. Law}
\altaffiliation{NASA Hubble Fellowship Program Sagan Fellow}
\affiliation{Department of Astronomy, University of Virginia, Charlottesville, VA 22904, USA}
\email{cjl8rd@virginia.edu}

\author[0000-0002-7058-7682]{Hideko Nomura}
\email{hideko.nomura@nao.ac.jp}
\affiliation{National Astronomical Observatory of Japan, 2-21-1 Osawa, Mitaka, Tokyo 181-8588, Japan}
\affiliation{Department of Astronomical Science, The Graduate University for Advanced Studies, SOKENDAI, 2-21-1 Osawa, Mitaka, Tokyo 181-8588, Japan}

\author[0000-0002-1199-9564,sname='Perez']{Laura Perez}
\email{lperez@das.uchile.cl}
\affiliation{Departamento de Astronomía, Universidad de Chile, Camino El Observatorio 1515, Las Condes, Santiago, Chile}

\author[0000-0003-4853-5736,sname='Rosotti']{Giovanni Rosotti}
\email{giovanni.rosotti@unimi.it}
\affiliation{Dipartimento di Fisica, Universit\`{a} degli Studi di Milano, Via Celoria 16, 20133 Milano, Italy}

\author[0000-0003-2993-5312]{Yuhito Shibaike}
\email{shibaike.y.aa@gmail.com}
\affiliation{Graduate School of Science and Engineering, Kagoshima University, 1–21–35 Korimoto, Kagoshima City, Kagoshima 890–0065, Japan}

\author[0000-0003-1534-5186]{Richard Teague}
\email{rteague@mit.edu}
\affiliation{Department of Earth, Atmospheric, and Planetary Sciences, Massachusetts Institute of Technology, Cambridge, MA 02139, USA}

\author[0000-0002-6034-2892]{Takashi Tsukagoshi}
\email{yyamato.as@gmail.com}
\affiliation{Faculty of Engineering, Ashikaga University, Ohmae 268-1, Ashikaga, Tochigi, 326-8558, Japan}

\author[0000-0003-4099-6941]{Yoshihide Yamato}
\email{yyamato.as@gmail.com}
\affiliation{RIKEN Pioneering Research Institute, 2-1 Hirosawa, Wako, Saitama 351-0198, Japan}


\begin{abstract}
Despite observational progress in planet formation, the stage in which planetesimals grow into planets remains poorly understood.
During this phase, protoplanets may develop gaseous envelopes that are warmer than the surrounding disk gas, potentially providing observable signatures through molecules otherwise depleted in cold regions.
In this Letter, we report the detection of the silicon sulfide isotopologues \isis\ and \jsis\ in the protoplanetary disk around PDS 66 (MP Mus) at a significance of ${\sim}5{-}6\sigma$, using the Atacama Large Millimeter/submillimeter Array.
These constitute the second and first detections of $\rm ^{28}SiS$ and $\rm ^{30}SiS$ in a protoplanetary disk, respectively.
The emission appears as a compact source at $r=60$ au in the southwestern region of the disk, unresolved with a ${\sim}0\farcs5$ beam, and shows a velocity consistent with Keplerian rotation, suggesting a protoplanetary origin. 
By modeling the line fluxes, we constrain the emitting radius to ${\sim}0.5{-}4$ au and estimate an SiS mass of $10^{22}{-}10^{23}$ g, corresponding to at least ${\sim}10\%$ of the silicon contained in local dust grains.
Because complete sublimation of a substantial fraction of dust grains by local processes is difficult to achieve, this result instead implies an accumulation of silicon from a larger region.
We propose that a circumplanetary envelope surrounding a low-mass protoplanet, where pebble accretion and subsequent sublimation of grains may enhance gaseus silicon abundance with respect to observable dust grains around it, can account for the observed characteristics.
\end{abstract}


\section{Introduction}

Planet formation is a process in which small dust grains grow into planets within protoplanetary disks \citep[e.g.,][]{2023ASPC..534..717D}.
In the initial stage, micron-sized dust grains undergo collisional growth and accumulate into relatively large particles, or pebbles.
These pebbles may further evolve into planetesimals, potentially through the streaming instability \citep{2005ApJ...620..459Y}.
The planetesimals then accrete other planetesimals or pebbles under the presence of aerodynamic drag from the disk gas, allowing them to grow to the sizes of terrestrial planets or the cores of giant planets \citep{2010A&A...520A..43O, 2012A&A...544A..32L}.
Once the cores become sufficiently massive, they can accrete disk gas through circumplanetary disks.
Recent observational studies have revealed that these evolutionary stages are indeed occurring.
Dust evolution in disks has been recognized as a variation in the continuum spectral index at millimeter wavelengths \citep[e.g.,][]{2014prpl.conf..339T, 2023ASPC..534..501M}.
Radial dust accumulation is a necessary condition for the streaming instability, and has been suggested by high resolution ALMA observations \citep[e.g.,][]{2018ApJ...869L..41A}.
Relatively massive protoplanets undergoing gas accretion have been directly detected \citep[e.g.,][]{2018A&A...617A..44K, 2019NatAs...3..749H}, together with their circumplanetary disks \citep[e.g.,][]{2021ApJ...916L...2B}. Additionally, gas kinematics traced by molecular/atomic lines \citep[e.g.,][]{2018ApJ...860L..13P, 2019Natur.574..378T, 2022ApJ...941L..24A} and spatially-localized, excess line emission \citep[e.g.,][]{2022ApJ...934L..20B, 2023A&A...669A..53B, 2023ApJ...952L..19L} also suggest the presence of embedded giant planets.

However, the evolutionary pathway from planetesimals to massive (${\sim}1\ M_{\rm Jup}$) protoplanets remains poorly constrained from an observational perspective. This is due at least in part to the fact that the mass of these nascent planets are much less than that of a Jupiter as they are initially forming/evolving.
It is therefore important to identify observational signatures that can probe this intermediate stage.
In the presence of disk gas, planetesimals or small protoplanets can acquire envelopes \citep[e.g.,][]{2016MNRAS.460.2853S}.
Such envelopes extend out to the Hill (or Bondi) radius, which reaches ${\sim}1$ au for an Earth-sized planet located at 100 au from the central star.
This implies that a circumplanetary envelope at the outer regions of disks can attain a size that can be probed with high-resolution observations.

Within such circumplanetary envelopes, significant heating processes may occur.
In the context of pebble accretion, pebbles accreted onto an envelope surrounding relatively massive cores (${\sim}1\ M_\oplus$) may evaporate before reaching the core, thereby limiting core growth \citep{2018A&A...611A..65B}.
The sublimated or thermally processed material may later recondense as the protoplanet cools or may be recycled into the protoplanetary disk \citep{2018A&A...611A..65B, 2023MNRAS.523.6186W}.
Although it remains uncertain which species can most effectively trace these evolutionary stages, promising candidates include molecules that are strongly depleted in cold disk environments but released or formed in relatively heated and/or shocked environments, such as silicon-bearing molecules \citep{2015ApJ...807....2C, 2023ApJ...952L..19L, 2024MNRAS.534.3576K, 2025A&A...693A..86C}.

In this Letter, we report the detection of two silicon sulfide (SiS) isotopologues, $\rm ^{28}Si^{32}S$ and $\rm ^{30}Si^{32}S$, at a distance of ${\sim}60$ au from the central star in the protoplanetary disk around PDS 66 (MP Mus).
The PDS 66 disk is among the nearest protoplanetary disks to Earth, at a distance of $d=97.9$ pc \citep{2021A&A...649A...1G}.
The central star is a K1V-type star \citep{2002AJ....124.1670M}, with estimated mass, luminosity, age, and mass accretion rate of $M_\star \simeq 1.28\ M_\odot$, $L_\star \simeq 1.2\ L_\odot$, 7–10 Myr, and $1.3\times10^{-10}\ M_\odot\ {\rm yr^{-1}}$, respectively \citep{2013ApJ...767..112I, 2023A&A...673A..77R, 2025ApJ...984L...8I}.
The disk has been imaged at infrared and millimeter wavelengths.
The near-infrared (H-band) image shows a gap-like structure at a radius of ${\sim}60$ au from the central star \citep{2016ApJ...818L..15W, 2018ApJ...863...44A}.
High-resolution (4 au) Atacama Large Millimeter/submillimeter Array (ALMA) Band 6 ($\lambda{\sim}1.3$ mm) observations reveal that the continuum disk extends to ${\sim}60$ au in radius with no detected substructure \citep{2023A&A...673A..77R}.
A lack of substantial substructures in high-resolution observations is relatively uncommon in large disks \citep{2023ASPC..534..423B}, although follow-up studies have revealed shallow substructures \citep{2025A&A...698A.165A, 2025NatAs...9.1176R}.
An upper limit of ${\sim}1\ M_{\rm Jup}$ on planets in the outer disk regions has been placed based on CO channel maps \citep{2025ApJ...984L..15P}.

The structure of this Letter is as follows; Section \ref{sec:res} describes the observations and results, Section \ref{sec:disc} presents the interpretations and discussion, and Section \ref{sec:sum} summarizes the conclusions.

\section{Observations and Results} \label{sec:res}

\subsection{Observations}
\begin{deluxetable}{l c c c c c c c c}
\tablecaption{Observed SiS isotopologue transitions\label{tab:sis_obs}}
\tablewidth{0pt}
\tablehead{
\colhead{Transition} &
\colhead{CDMS Freq.} &
\colhead{JPL Freq.} &
\colhead{$E_{\rm up}$} &
\colhead{$g_u A_{ul}$} &
\colhead{Beam size (P.A.)} &
\colhead{RMS} &
\colhead{Int. Flux} \\
\colhead{} &
\colhead{(GHz)} &
\colhead{(GHz)} &
\colhead{(K)} &
\colhead{($10^{-2}\ \rm s^{-1}$)} &
\colhead{} &
\colhead{($\rm mJy\ beam^{-1}$)} &
\colhead{($\rm mJy\ \kms$)}
}
\startdata
$^{28}$SiS $J{=}16{-}15$ & $290.3808$ & $290.3803$ & $118$ & $1.37$ & $0\farcs62\times0\farcs42\ (35^\circ)$ & $1.1$ & $21\pm4$ \\
$^{30}$SiS $J{=}18{-}17$ & $315.0625$ & $315.0659$ & $144$ & $1.97$ & $0\farcs57\times0\farcs41\ (-6^\circ)$ & $1.7$ & $12\pm2$ \\
\enddata
\tablecomments{RMS values are measured for images with a channel width of $1.25\ \kms$.}
\end{deluxetable}

We observed the ${\rm ^{28}Si^{32}S}\ J{=}16{-}15$ and ${\rm ^{30}Si^{32}S}\ J{=}18{-}17$ pure rotational transitions toward the PDS 66 disk.
Hereafter, we omit the atomic number 32 for the main sulfur isotopologue.
The two lines were observed by different ALMA programs (${\rm ^{28}SiS}$: \#2023.1.00334.S, PI: L. Trapman; ${\rm ^{30}SiS}$: \#2023.1.00525.S, PI: T.C. Yoshida).

$\rm ^{28}Si$ is the most abundant stable isotope of silicon, while $\rm ^{30}Si$ is the least abundant.
The solar isotope ratio of $\rm ^{28}Si/^{30}Si$ is estimated to be $30.1$ \citep{2021A&A...653A.141A}.
Table \ref{tab:sis_obs} lists the spectroscopic parameters of the observed lines.
We provide two rest frequencies from the CDMS \citep{2005JMoSt.742..215M} and JPL \citep{1998JQSRT..60..883P} databases.
The CDMS and JPL entries adopt different molecular constants based on \citet{2007PCCP....9.1579M} and \citet{1974JPCRD...3..609L}, respectively.
Unless otherwise noted, we use the constants on the CDMS catalogue in this work.

Details of the observations and data reduction are described in Appendix \ref{app:obs}.
In summary, we produced two image cubes for the ${\rm ^{28}SiS}\ J{=}16{-}15$ transition, one with a narrow channel width of $\Delta v = 1.25\ \kms$ and another with a wider channel width of $5.0\ \kms$ to maximize the S/N of a potentially broad line.
For the ${\rm ^{30}SiS}\ J{=}18{-}17$ transition, only the narrow channel-width cube was generated.
The resulting beam sizes and noise levels are summarized in Table \ref{tab:sis_obs}.
Channel maps of the narrow-channel data cubes are presented in Appendix \ref{app:obs}.

\subsection{Detection of SiS isotopologues} \label{sec:res1}

\begin{figure*}
\centering
\includegraphics[width=0.9\textwidth]{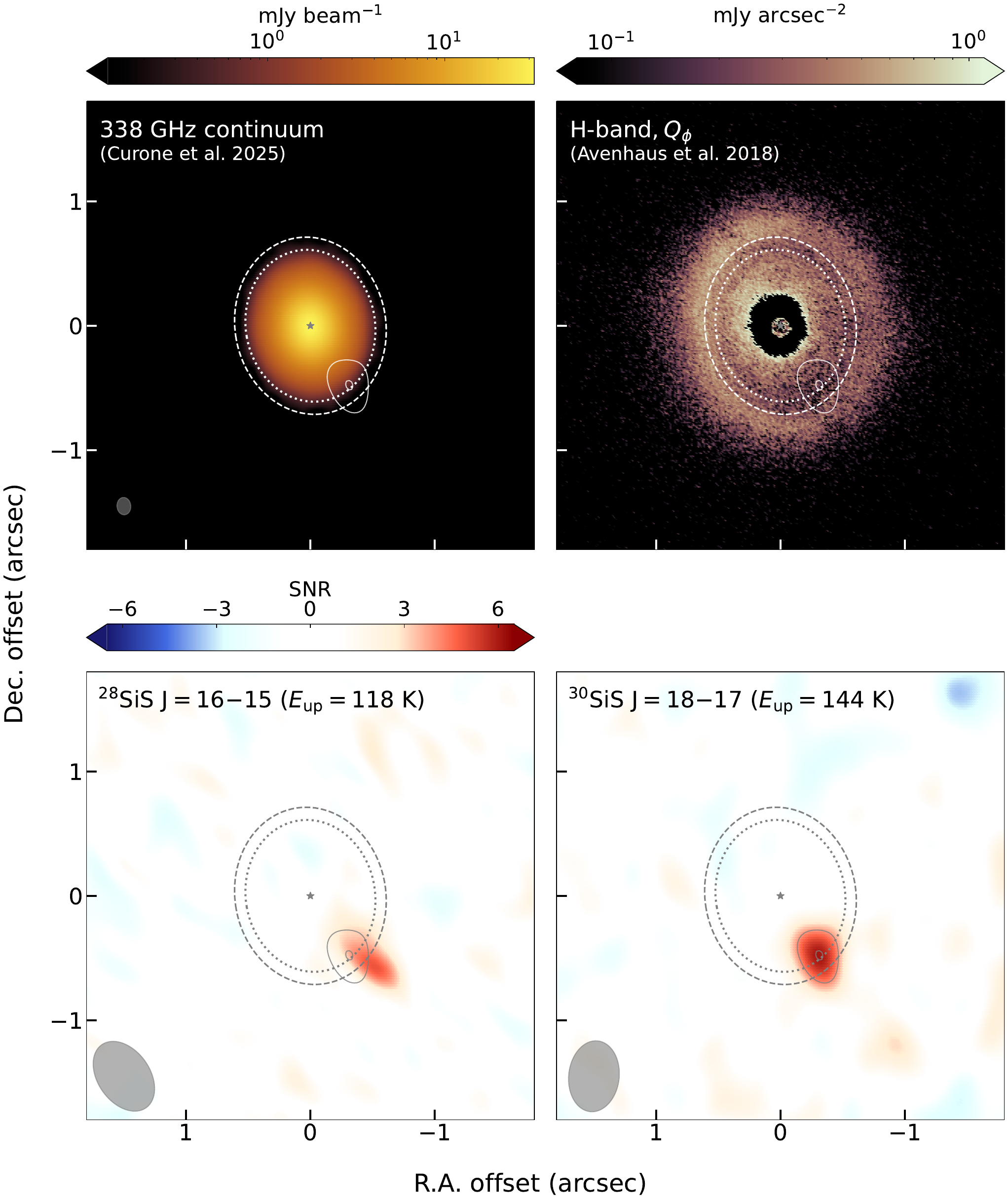}
\caption{Upper panels: 338 GHz continuum image \citep{2025ApJ...984L...9C} and $Q_{\phi}$ image in the H band \citep{2018ApJ...863...44A}. Bottom panels: S/N maps of the \isis\ and \jsis\ lines at channels that cover $v_{\rm LSR} \simeq 2.5-7.5\ \kms$ and $3.1-4.4\ \kms$, where the emission peaks. Note that there might be uncertainty in the rest frequency adopted here (Appendix \ref{app:freq}).
In all panels, the dotted and dashed ellipses indicate radii of $r=60$ and $70$ au, corresponding roughly to the edge of the mm-continuum disk \citep{2023A&A...673A..77R, 2025ApJ...984L...9C}. Contours show the 3 and $6\sigma$ levels of the \jsis\ line, and the apparent shift between two lines can be explained by statistical error of astrometry (Section \ref{sec:res1}).  The star symbol marks the position of the central star, and the gray ellipse in the lower left corner represents the synthesized beam (except for the H-band image).}
\label{fig:mom0}
\end{figure*}

We detected significant emission in both image cubes.
In the $v_{\rm LSR}=5\ \kms$ channel of the wide-channel \isis\ cube, we found ${\sim}5\sigma$ emission in the southwestern part of the disk.
In the $v_{\rm LSR}=3.75\ \kms$ channel of the \jsis\ cube, a ${\sim}6\sigma$ signal was detected at a similar position.
This is the first detection of ${\rm ^{30}SiS}$ in protoplanetary disks.
Figure \ref{fig:mom0} shows the signal-to-noise ratio (S/N) channel maps for both lines, and Appendix \ref{app:hist} presents the pixel-value histograms.
The sources are spatially unresolved and have position angles consistent with the synthesized beam, indicating that the emission originates from a point source at the current spatial resolution (${\sim}0\farcs5$).
The upper state energies of those lines are $>100$ K, which is significantly higher than the expected temperature at this region from the stellar luminocity (${\sim}17$ K; Appendix \ref{app:caveats}).

After we specified the disk center by fitting a Gaussian to the corresponding continuum images, we measured the relative position of the emission peaks with respect to the disk center as $({\rm \Delta R.A.,\ \Delta Dec.})=(-0\farcs51,-0\farcs55)$ and $(-0\farcs31,-0\farcs47)$ for the \isis\ and \jsis\ lines, respectively.
The separation between the two positions is ${\sim}0\farcs21$.
According to the ALMA Technical Handbook \citep{cortes_2025_14933753}, the 1$\sigma$ relative astrometric accuracy for sources with these S/Ns is ${\sim}0\farcs17$, indicating that the apparent offset is consistent with the statistical error.

Figure \ref{fig:mom0} also compares the SiS emission with the 338 GHz continuum
\footnote{The continuum image was obtained from the exoALMA repository on the Harvard Dataverse \citep{DVN/DMM54B_2025}. Note that the image frequency (338 GHz) differs slightly from the one used in the original paper (332 GHz).} \citep{2025ApJ...984L...9C} and H-band polarized-intensity image \citep{2018ApJ...863...44A}.
Dotted ellipses indicate $r=60$ au, corresponding to the outer radius of the mm-continuum disk \citep{2023A&A...673A..77R}.
The detected point source lies near the disk edge and within the infrared gap.
An azimuthally averaged radial profile analysis of the continuum disk suggests a more extended structure up to $r{\sim}70$ au \citep{2025ApJ...984L...9C}, which is shown in dashed ellipses in Figure \ref{fig:mom0} and slightly beyond the SiS emission region.
In addition, interestingly, the SiS emission is co-located with the azimuthal brightness drop in the H and K1-band image with Gemini Planet Imager \citep{2016ApJ...818L..15W}.
They attributed this to a shadow cast by material in the inner region or cold stellar spots.
Note that, however, \citet{2018ApJ...863...44A} did not find this structure in their H-band image with VLT/SPHERE.
\begin{figure}
\centering
\includegraphics[width=0.5\textwidth]{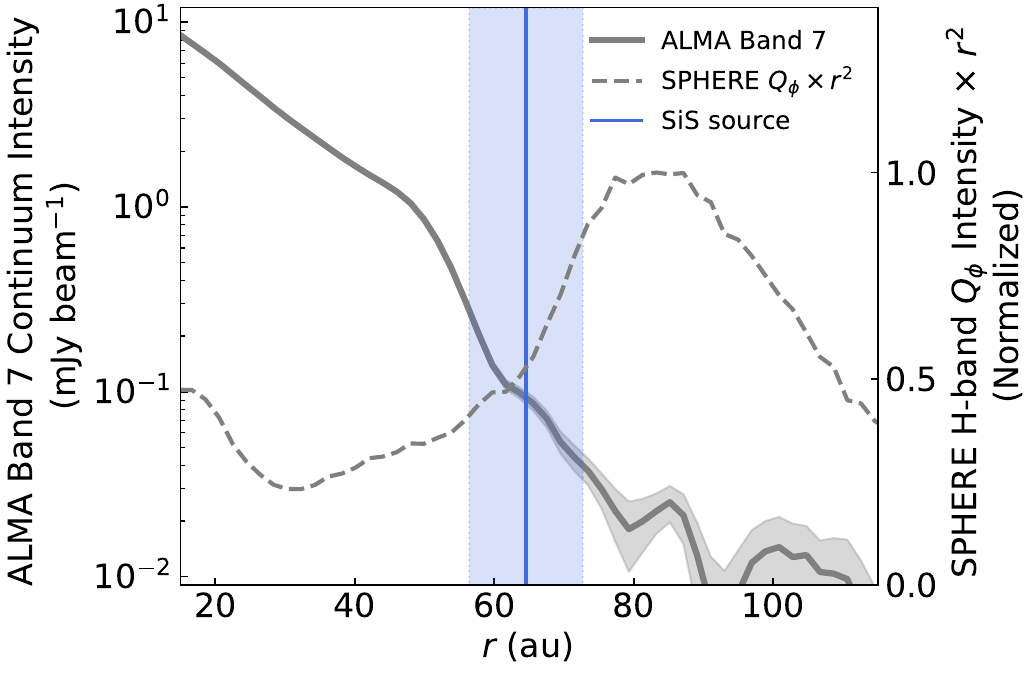}
\caption{
Radial location of the SiS source with the radial profiles of the 332 GHz continuum and H-band $Q_\phi$ emission multiplied by $r^2$ \citep{2025ApJ...984L...9C, 2018ApJ...863...44A}.
Note that the SiS source location is an averaged value of the two isotopologues weighted by their astrometric uncertainties.
The blue-shaded region shows the $1\sigma$ uncertainty of the averaged location.}
\label{fig:radprof}
\end{figure}
To summarize the comparison of radial locations, we also plot the deprojected radial profiles of the 338 GHz continuum \citep{2025ApJ...984L...9C} and H-band polarized-intensity (weighted by $r^2$) \citep{2018ApJ...863...44A} with the radial location of the SiS source obtained by weighted average of the positions of the two isotopologues (Figure \ref{fig:radprof}).

\begin{figure}
\centering
\includegraphics[width=0.4\textwidth]{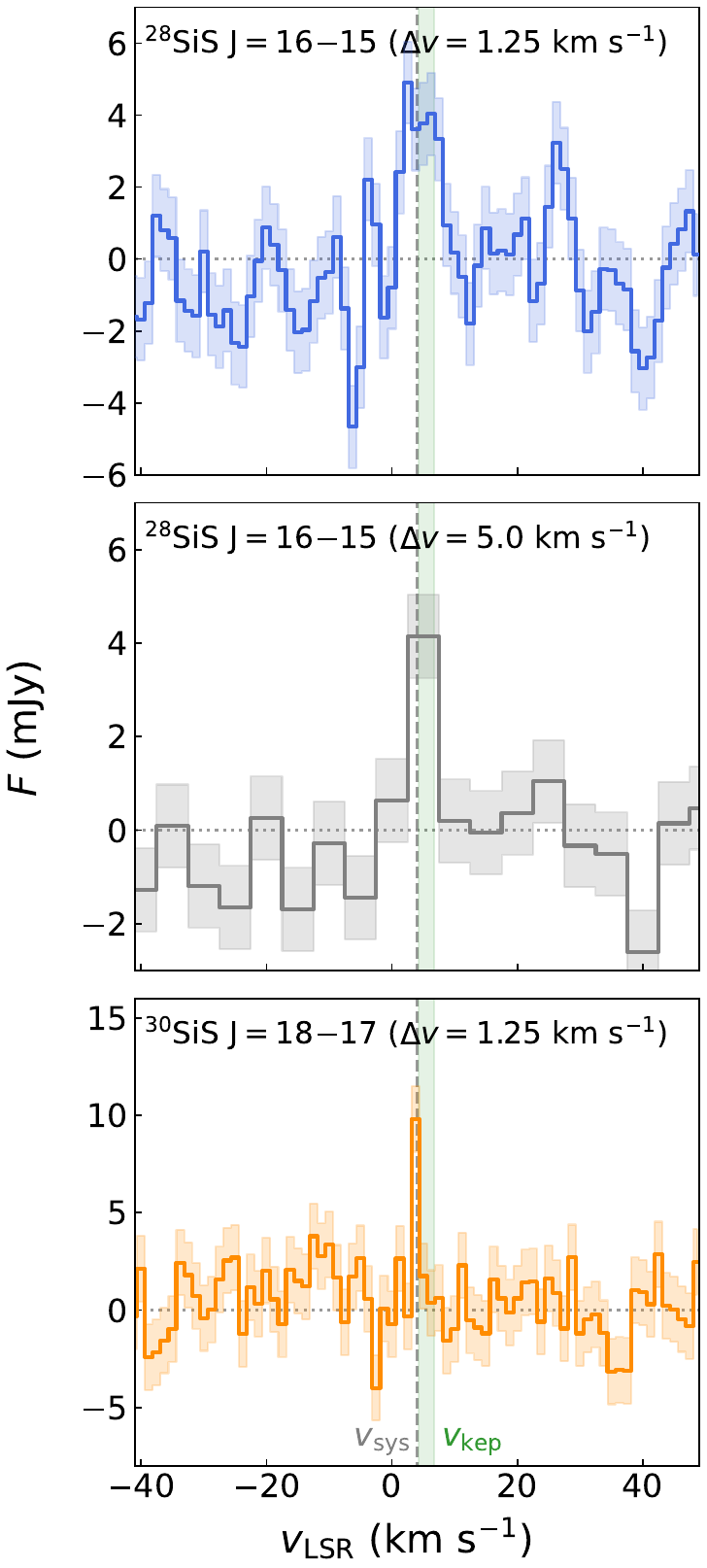}
\caption{Spectra of the \isis\ and \jsis\ lines extracted at the peak S/N positions (Figure \ref{fig:mom0}). The top and bottom panels show spectra from the narrow channel-width cubes, while the middle panel corresponds to the wide-channel \isis\ cube. The shaded regions indicate $1\sigma$ noise levels. The vertical dashed line marks the systemic velocity, and the green band represents the range of projected Keplerian velocities within one beam.}
\label{fig:spec}
\end{figure}

Figure \ref{fig:spec} presents the spectra at the pixels where the emission peaks.
In the narrow-channel \isis\ cube, ${\sim}3{-}4\sigma$ signals appear over five channels around $v_{\rm LSR}\sim5\ \kms$, implying a line width of ${\sim}5\ \kms$ (FWHM).
This corresponds to the ${\sim}5\sigma$ detection in the wide-channel cube shown in the second panel.
For the \jsis\ line, the signal appears only at $v_{\rm LSR}=3.75\ \kms$ and is not spectrally resolved at a velocity resolution of ${\sim}1\ \kms$.
Although the signal-to-noise ratio is low, we found that the two lines have potentially different line widths, which we discuss in Appendix \ref{app:caveats}.
The integrated flux densities are $21\pm4$ and $12\pm2\ {\rm mJy\ \kms}$ for the \isis\ and \jsis\ lines, respectively.

The systemic velocity of PDS 66 has been measured as $v_{\rm LSR}\simeq3.98\ \kms$ from CO observations \citep{2023A&A...673A..77R, 2025ApJ...984L...8I}.
The line-of-sight Keplerian velocity within the beam of the compact source ranges between $v_{\rm LSR}=4$ and $6\ \kms$.
Thus, the \isis\ source is consistent with the local Keplerian velocity, while the \jsis\ source appears slightly offset. However, this is likely within the uncertainty of the rest frequency (Appendix \ref{app:freq}).
Overall, the SiS emission is both astrometrically and spectrally consistent with the PDS 66 system, strongly suggesting an association with the disk.
While consistent proper motion between the emission and the disk (or the star itself) would further strengthen this scenario, the two observations were conducted only seven months apart. The expected stellar proper motion during this window is $\sim 38$ mas \citep{2021A&A...649A...1G}, which is insufficient to provide a definitive constraint. Nevertheless, a background or foreground origin is unlikely for the following reasons.
Potential sources with previous SiS detections, such as AGB stars \citep[e.g.,][]{2011ApJS..193...17P, 2019MNRAS.484..494D}, massive protostellar environments \citep[e.g.,][]{2011A&A...528A..26T, 2020ApJ...900L...2T, 2023ApJ...942...66G}, or protostellar outflows \citep[e.g.,][]{2017MNRAS.470L..16P}, typically exhibit strong continuum emission around them, which is not observed here.
For example, AGB stars at 130–2200 pc have millimeter fluxes of 3–70 mJy \citep{2025arXiv250400517D}, and the carbon star IRC+10216 at 135 pc, where SiS isotopologues are detected, shows ${\sim}650$ mJy \citep{2011ApJS..193...17P}.
The detection of $\rm ^{28}SiS$ in the protoplanetary disk around HD 169142 \citep{2023ApJ...952L..19L} further supports the interpretation that the SiS emission originates in-situ from the PDS 66 system.

\subsection{Derived physical quantities} \label{sec:res2}

\begin{figure}
\centering
\includegraphics[width=0.45\textwidth]{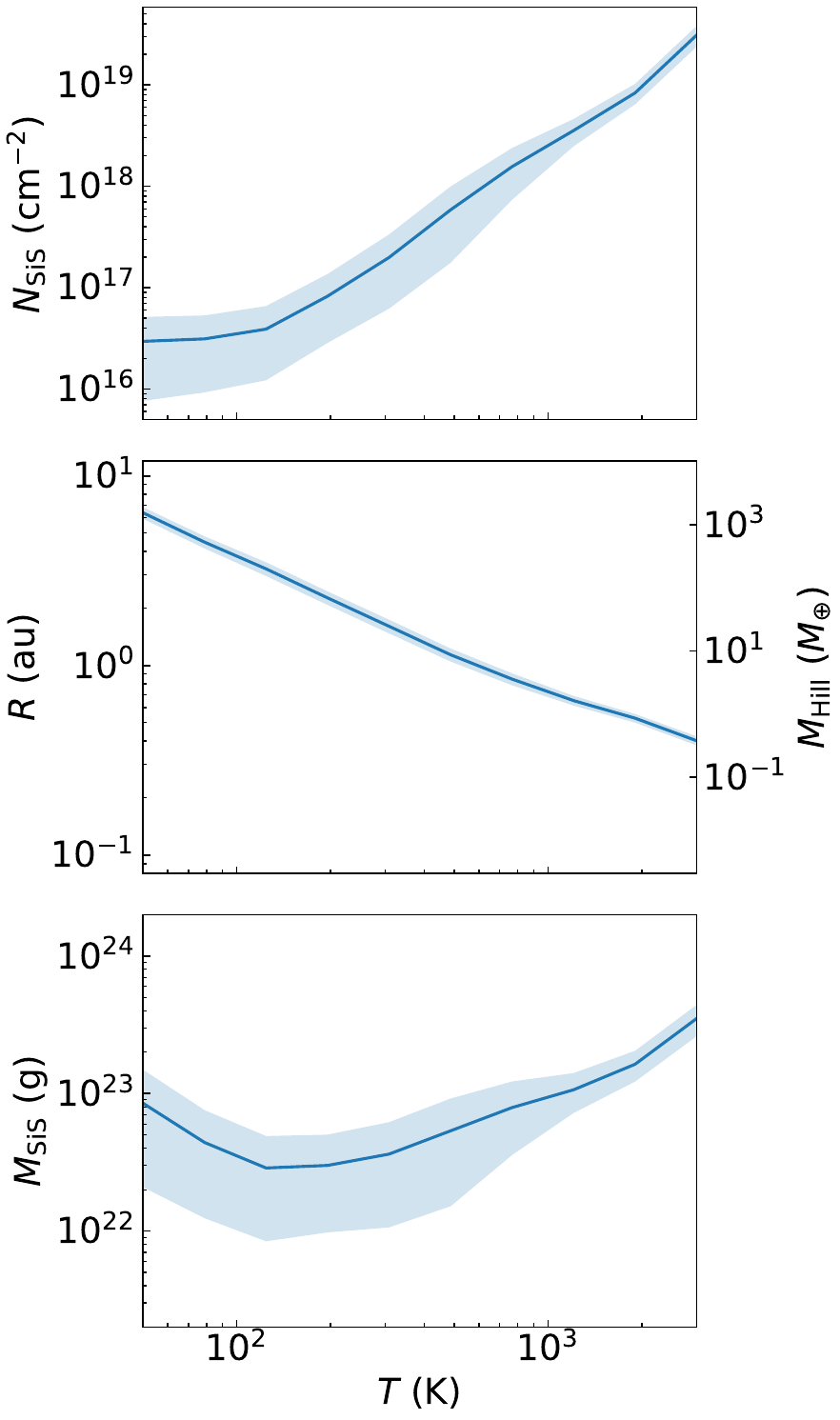}
\caption{Derived SiS column densities ($N_{\rm SiS}$) and emitting radii ($R$) as a function of the assumed temperature. The bottom panel shows the SiS mass ($M_{\rm SiS}$). For the middle panel ($R$), we also plot the planet's mass assuming that the emitting radius equals to the Hill radius (see Section \ref{sec:pa}). } 
\label{fig:quantities}
\end{figure}

The detection of the rare isotopologue $\rm ^{30}SiS$ indicates that the total SiS mass is substantial.
To quantify the physical conditions of the source, we modeled the integrated flux densities of the detected lines under the assumptions of local thermal equilibrium, a homogeneous cylindrical emitting region with radius $R$, and an isotopologue ratio $\rm ^{28}SiS/^{30}SiS = 30$ \citep{2021A&A...653A.141A}.
Spectroscopic parameters were taken from the CDMS database \citep{2005JMoSt.742..215M, 2007PCCP....9.1579M}, and integrated line fluxes were computed over a grid of temperatures $T$, SiS column densities $N_{\rm SiS}$, and emitting radii $R$.

In our model, the integrated flux of each transition was computed as
\begin{equation}
F_{i} = \pi R_a^2 \int_{-\infty}^{\infty} B_{\nu_i}(T)\,\left(1 - e^{-\tau_i(v)}\right)\,dv,
\end{equation}
where $i$ denotes each line, $R_a$ is the apparent radius corresponding to a physical radius $R$ at a distance of 98 pc, $\nu_i$ is the frequency of the transition, $B_{\nu_i}(T)$ is the Planck function, and $\tau_i(v)$ is the optical depth as a function of velocity offset from the line center.
We compared $F_{i}$ with the measured values and identified combinations of $(N_{\rm SiS}, R, T)$ that reproduce both integrated fluxes within the one sigma uncertainties.

Parameter sets consistent with the observations are shown in Figure \ref{fig:quantities}.
For $T = 100$–$2000$ K, the SiS column density increases from ${\sim}10^{16}$ to $10^{19}\ {\rm cm^{-2}}$, while the emitting radius decreases from ${\sim}4$ to $0.5$ au.
The corresponding SiS mass is $M_{\rm SiS} \sim 10^{22}$–$10^{23}\ {\rm g}$.
As expected from the observed flux ratio, the \isis\ line is highly optically thick, with peak optical depths of $30$–$200$, which constrains the emitting radius.
In contrast, the \jsis\ line is roughly $30$ times less optically thick and is therefore still sensitive to the column density.
Caveats associated with these calculations are discussed in Appendix \ref{app:caveats}.

\section{Discussion} \label{sec:disc}

\subsection{Dust Mass Infered from SiS and Continuum}
In general, silicon-bearing molecules are highly locked in the solid phase of protoplanetary disks, namely, dust grains.
Therefore, the detection of SiS indicates a dynamically active environment in the disk, even a possible protoplanetary origin.
The main reservoir of silicon in dust grains is expected to be silicates forming the grain cores, such as $\rm Mg_2SiO_4$, which may constitute ${\sim}33\%$ of the total dust mass \citep[e.g.,][]{1994ApJ...421..615P, 2018ApJ...869L..45B}.
A few percent of the silicon may also reside in the grain mantles, potentially as SiO, as suggested by observations of protostellar outflows \citep{2008A&A...490..695G}.
Releasing silicon from the grain core requires significant heating ($>1000$ K) or strong shocks, although the condition is much less stringent for grain mantles.
Once released into the gas phase, gas-phase reactions with S-bearing molecules are proposed to form SiS molecules \citep{2018CPL...695...87R, 2020MNRAS.493..299P, 2021ApJ...920...37M, 2018ApJ...862...38Z, 2024A&A...687A.149M, 2024ApJ...971..101F}.

In any case, the silicon detected in SiS should originate from dust grains.
Assuming that the fraction of silicon released from one dust grain is $\chi$ ($\chi\simeq1$ if the entire grain is sublimated, while $\chi \lesssim 0.1$ if only the mantle contributes), the corresponding dust mass needed to account for silicon in SiS can be estimated as
\begin{eqnarray} \label{eq:dsis}
    M_{d,\ \rm SiS} &\bf{\sim}& 15 \chi^{-1} M_{\rm SiS} \\ 
    &\sim& (10^{-5} - 10^{-4})\chi^{-1}\ M_\oplus \\
    &\sim& (0.1 - 1) \chi^{-1}\ M_{\rm Ceres}
\end{eqnarray}
for $T=100$–$2000$ K using the results in Section \ref{sec:res2}.
Here, the factor of ${\sim}15$ is the mass ratio of the dust mass to Si, assuming that Si is dominated by silicate that is five times heavier than Si and has a mass fraction of ${\sim}33\%$ of one grain.
This indicates that even if the silicon originates from sublimation of dust grain cores, at least $10^{-5} \ M_\oplus$ of dust is required.
Moreover, if the silicon comes only from grain mantles, an even larger dust mass is necessary.

\begin{figure}
\centering
\includegraphics[width=0.45\textwidth]{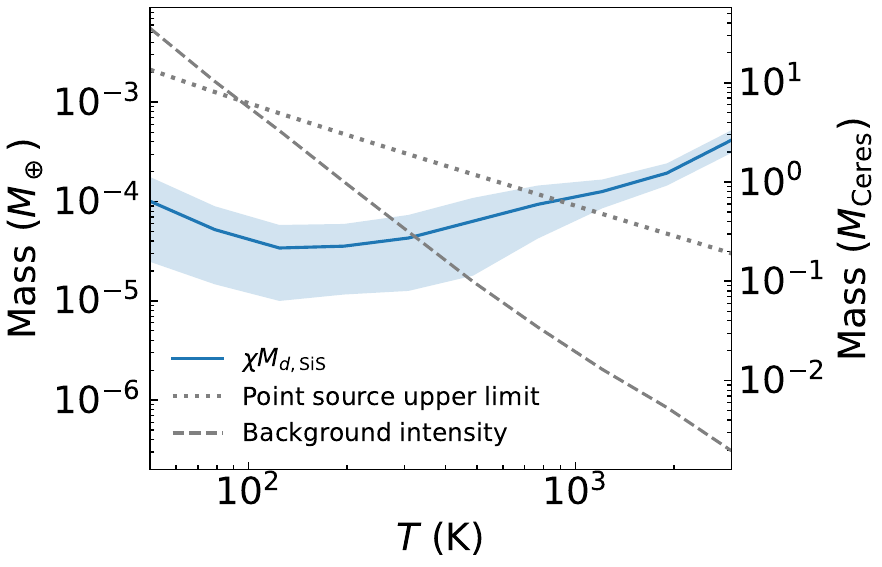}
\caption{Dust mass required to supply gas-phase SiS compared with dust masses inferred from background continuum intensity and the upper limit from a potential point source. $\chi$ is the fraction of silicon released from one dust grain.}
\label{fig:masscomp}
\end{figure}

We compare the dust mass needed to account for the SiS, $M_{d,\ \rm SiS}$, with continuum observations.
First, we assume dust grains are uniformly distributed within the SiS-emitting region, with a radius of ${\sim}0.4$–$5$ au, and the surface brightness matches that of the surrounding azimuthal region.
The 332 GHz continuum intensity at $r = 60$ au is estimated as $I_{\rm cont} \sim 6 \times 10^8\ {\rm Jy\ Sr^{-1}}$ by \citet{2025ApJ...984L...9C} using {\tt frankenstein} fitting \citep{2020MNRAS.495.3209J}.
Assuming the millimeter opacity of a typical protoplanetary disk dust mixture, $\kappa=3.5\ {\rm cm^2\ g^{-1}}$ \citep{2018ApJ...869L..45B},  and a temperature equal to that of SiS, the dust mass within radius $R$ is 
\begin{equation}
    \frac{ \pi R^2 I_{\rm cont}}{ \kappa B_\nu (T) } \sim10^{-6} - 10^{-3}\ M_\oplus,
\end{equation}
depending on the assumed temperature ($T=10^2-10^3$ K).

However, the azimuthally and radially localized SiS emission may also suggest a corresponding localized dust excess.
Although previous sensitive millimeter-continuum ALMA observations reveal no point-source excess at the SiS location \citep{2023A&A...673A..77R, 2025ApJ...984L...9C, 2025NatAs...9.1176R}, we estimate the $3\sigma$ upper limit of the excess dust mass using the continuum noise level \citep[$\sigma\sim25\ {\rm \mu Jy}$ at 332 GHz;][]{2025ApJ...984L...9C} via 
\begin{equation}
\frac{3\sigma d^2}{\kappa B_\nu(T)} \sim 10^{-4} - 10^{-3}\ M_\oplus.
\end{equation}
Figure \ref{fig:masscomp} shows these dust mass estimates from the background intensity and point-source upper limit as a function of assumed temperature.
Interestingly, the dust masses inferred from continuum observations are roughly comparable to the dust mass required to supply the observed SiS.
At $T{\sim}100$ K, at least ${\sim}10\%$ (and likely more) of the local dust grains must have sublimated.
At $T{>}1000$ K, the silicon content in the local dust grains under the point-source assumption reaches the total SiS silicon, implying that all dust grains in this region may have sublimated.
In summary, we find that if the SiS originates from a local process in a closed system, the silicate dust cores at this location must have been almost completely sublimated.

Alternatively, silicon may have been accumulated at this position relative to the amount of dust grains. 
In this case, release from grain mantles alone could also suffice if the degree of accumulation is substantial.
Such an enhancement can be divided into two non-exclusive possibilities: silicon could be selectively concentrated at the location, and/or the dust grains traced at millimeter wavelengths could be selectively depleted.
The latter case would correpond to conversion of dust grains to larger bodies such as planetesimals.
It is noteworthy that although our discussion has so far focused on silicon rather than sulfur, a gas enriched in silicon would be also expected to be sulfur-rich because sulfur is generally more volatile than silicon. Indeed, S-bearing molecules have been routinely detected in warm Herbig disks while Si-bearing molecules are not \citep[e.g.,][]{2025ApJ...989...30Z, 2025arXiv251201731B}.
In the following section, we consider four phenomenon that are potentially consistent with the above discussion.

\subsection{Possible Origin of SiS}

\subsubsection{Planetary-driven outflow}
Another detection of SiS in a protoplanetary disk has been reported in HD 169142 \citep{2023ApJ...952L..19L}.
However, its characteristics differ from PDS 66; the emission is extended rather than a point source, and it is spectrally offset by ${\sim}6\ \kms$ from the local Keplerian rotation.
In that case, the emission was attributed to a polar outflow driven by the massive protoplanet candidate HD 169142 b \citep[e.g.,][]{1998ApJ...508..707Q, 2022ApJ...941L..24A, 2023MNRAS.522L..51H, 2024ApJ...971L..15Y, 2024MNRAS.534.3576K}.
SiS could be released directly from the dust grains by strong shocks or formed via gas-phase reactions of Si and other species after release from the grain mantles \citep[e.g., ][]{2018CPL...695...87R, 2020MNRAS.493..299P, 2021ApJ...920...37M, 2024A&A...687A.149M, 2024ApJ...971..101F}.

Nevertheless, a planet-driven outflow is unlikely to account for the emission in PDS 66.
The detected velocity is consistent with the local Keplerian speed, and the emission is compact, with an inferred spatial scale of ${\sim}0.4$–$5$ au.
These characteristics suggest that the SiS originates from a component that is gravitationally bound to the potential planet rather than an outflow.
It also remains unclear whether an outflow could supply the large local fraction of silicon required relative to the silicon contained in the local dust grains.

\subsubsection{Circumplanetary Disk}
Another possibility is that the SiS emission originates from a hot region in a circumplanetary disk, where dust grains can be thermally evaporated.
However, \citet{2024A&A...687A.166S} showed that even in the circumplanetary disk around the giant planet PDS 70 c, temperatures exceeding ${\sim}10^3$ K, which is needed to thermally sublimate silicate, occur only in the very inner region, within ${\sim}0.1$ au.
This implies that most silicate in the disk should remain solid, which is inconsistent with the observations.
Still, as an analogy of young stellar objects \citep[e.g.,][]{2014ApJ...795...61B}, a potential outburst of the protoplanet could sufficiently heat the grains across large regions of the circumplanetary disk.
A lower temperature could also release the silicon from the dust mantles, it is unknown if there is any mechanism that can selectively deplete dust grains or enhance gas-phase silicon.

\subsubsection{Transient events}
Some transient events could lead to the sublimation of dust grains or larger bodies.
For example, giant impacts can vaporize a substantial amount of material \citep[e.g.,][]{2017JGRE..122..950L}, which could be consistent with the mass fractions as such giant bodies cannot be observed.
However, the duration of such events would be typically very short, so the probability of observing one by chance is low, making this scenario less likely than the others.
Future observations could test this possibility because the such emission should show time variability.

\subsubsection{Circumplanetary Envelope} \label{sec:pa}

A circumplanetary envelope may form around a relatively low-mass planet, with a typical spatial scale set by the Hill (or Bondi) radius.
During this stage, pebble accretion, a promising process for forming planetary cores, can operate \citep[e.g.,][]{2012A&A...544A..32L}.
If the protoplanet is relatively massive (${\sim}1\ M_\oplus$), rocky pebbles cannot reach the core but instead sublimate within the envelope, a process referred to as ``ablation'' \citep[e.g.,][]{2018A&A...611A..65B}.
This process limits direct core growth and enriches the envelope with silicate vapor \citep[e.g.,][]{2018A&A...611A..65B, 2023A&A...677A.181S, 2024A&A...683A.217S}.
For example, \citet{2024A&A...683A.217S} found that envelopes become saturated by SiO vapor for planet masses exceeding $0.6\ M_\oplus$, although their calculations assumed equilibrium chemistry and a protoplanet at 1 au.
We note that silicon contained in grain mantles has not been considered in this context, and such silicon may be released more easily into the gas phase at lower temperatures.

In the middle panel of Figure \ref{fig:quantities}, we plot the planet's mass assuming that the infered emitting radius equals to the Hill radius: $M_{\rm Hill} = 3 ( R / 60 {\rm\ au} )^3 M_\star $.
This suggests that the emitting radius is broadly consistent with the Hill radius of a $0.1$–$10\ M_\oplus$ planet at $r=60$ au, $0.26$–$1.2$ au.
$M_{d,\ {\rm SiS}}$ (Eq. \ref{eq:dsis}) corresponds to $ (10^{-6} - 10^{-3}) \chi^{-1}$ of these masses.
These low planet masses may also be consistent with the substructure-less disk \citep{2023A&A...673A..77R, 2025ApJ...984L..15P}, although there is a gap in the infrared image \citep{2018ApJ...863...44A}.
However, note that the masses would be lower than a typical range to exhibit an observable gap in scattered light images \citep{2017ApJ...835..146D}, potentially implying a different origin of the gap, such as a shadow \citep{2023A&A...673A..77R}.

We estimate the pebble accretion rate at this location.
The pebble accretion rate in the two dimentional regime can be roughly estimated as
\begin{equation} \label{eq:Macc}
\dot{M}_{\rm PA} \sim \sqrt{ G M_p {\rm St} } H_g r^2 \Sigma_d
\end{equation}
where $H_g$ is the gas scale height, $\Sigma_d$ is the dust surface density, $M_p$ is the planet mass, and ${\rm St}$ is the Stokes number \citep[e.g.,][]{2024arXiv241114643O}.
We adopt Eq. (\ref{eq:Hg}) for $H_g$ and assume ${\rm St} = 0.1$ for simplicity.
\citet{2025ApJ...984L...9C} measured the 332 GHz continuum intensity at $r = 60$ au (just outside the disk edge) to be $I_{\rm cont} \sim 6 \times 10^8\ {\rm Jy\ Sr^{-1}}$ using {\tt frankenstein} fitting \citep{2020MNRAS.495.3209J}.
Assuming a dust opacity of $\kappa = 3.5\ {\rm cm^2\ g^{-1}}$ \citep{2018ApJ...869L..45B} and optically thin emission, the dust surface density is estimated as
\begin{equation} \label{eq:sig_d}
\Sigma_d = \frac{I_{\rm cont}}{\kappa B_\nu(T_{\rm mid})} \sim 5 \times 10^{-3}\ {\rm g\ cm^{-2}}.
\end{equation}
Using Eq. (\ref{eq:Macc}), the pebble accretion rate for $M_p = 0.1-10\ M_\oplus$ is estimated to be $\dot{M}_{\rm PA} \sim 4 \times 10^{-7} - 4 \times 10^{-6}\ M_\oplus\ {\rm yr^{-1}}$.

This rate implies that the dust mass of $(10^{-5} - 10^{-4})\chi^{-1}\ M_\oplus$ can be supplied within ${\sim} ( 1 - 100 ) \chi^{-1} $ yr by pebble accretion, far shorter than not only the typical disk lifetime but also the orbital timescale.
We also note that planetesimal accretion onto the envelope may contribute to silicate vaporization \citep{2019ApJ...871..127V}.

The detailed mechanisms remain uncertain and are beyond the scope of this Letter.
Nevertheless, the observed SiS mass can be explained by the pebble accretion rate and an internal process in which only a very small fraction of the accreted dust grains are processed and their silicon is retained in the envelope.

\subsection{Consistency with non-detections of CO and SiO}

The SiS emission might suggest potential detections of other molecular lines.
For example, CO is a well-known tracer of molecular gas and has been observed by the exoALMA project \citep{2025ApJ...984L...6T}.
However, the CO channel maps presented by  \citet{2025ApJ...984L..15P} do not show a counter part of the SiS emission.
This is naturally explained if this emission is being absorbed by the optically thick protoplanetary disk gas above it, even if CO originates from the same source as SiS.
We note that this would not be the case if the two CO-emitting layers could be clearly distinguished in channel maps \citep[e.g.,][]{2022ApJ...934L..20B}, which is, however, not the case for the PDS 66 disk \citep{2025ApJ...984L..15P, 2025ApJ...984L..10G}.

SiO lines are another potential tracer, as they are sensitive to regions with strong shocks or heating \citep[e.g.,][]{2008A&A...490..695G, 2025Natur.643..649M}.
We searched the ALMA archive for all SiO $v=0$ transitions with velocity resolution better than ${\sim}20\ \kms$ and found that project ID \# 2017.1.01687.S (PI: A. Ribas) observed the SiO $J{=}3{-}2$ line ($E_{\rm up} = 13$ K) at 130.2687 GHz.
The observations were published by \citet{2023A&A...673A..77R}, and we refer to their paper for the observational details.
We obtained a pipeline-calibrated dataset from the National Radio Astronomy Observatory archive and concatenated the provided measurement sets.
After subtracting the continuum emission, we created image cubes with velocity widths of $1.25\ \kms$ and $5\ \kms$ using natural weighting.
No significant emission was detected, and we place a $3\sigma$ upper limit on the integrated intensity of $3.8\ {\rm mJy\ \kms}$ based on the narrow channel cube.
Assuming the same $R$ and $T$ for SiO as for SiS, the corresponding upper limit on the SiO column density is estimated to be $10^{15}-10^{18}\ {\rm g\ cm^{-2}}$, resulting in a lower limit of the SiS/SiO ratio of 5–60.
The high SiS/SiO ratio may reflect that SiS formation routes such as $\rm Si + SH \rightarrow SiS + H$ \citep{2021ApJ...920...37M} efficiently works.

Meanwhile, in AGB stars, where both SiS and SiO are routinely detected, the SiS/SiO ratio ranges from 0.05–20 \citep{2007A&A...473..871S}.
Notably, carbon stars typically have SiS/SiO $>1$, while O-rich M-type stars have SiS/SiO $<1$, reflecting the influence of the photospheric $\rm C/O$ ratio \citep{2006A&A...456.1001C}.
Although the chemistry in AGB stars may not directly apply to the PDS 66 system, the high SiS/SiO ratio may suggest $\rm C/O > 1$.

In protoplanetary disks, the C/O ratio is often higher than unity \citep[e.g.,][]{2016ApJ...831..101B, 2021ApJS..257....7B}.
Indeed, \citet{2023A&A...673A..77R} detected hydrocarbons in the PDS 66 disk, suggesting an enhaned C/O, although the exact value is currently unclear.
A likely mechanism to enhance C/O is removal of water ice by radial drift of icy grains \citep{2020ApJ...899..134K}.
In the context of the PDS 66 disk, the lack of substructure suggests active radial drift \citep{2023A&A...673A..77R, 2025A&A...698A.165A}, providing a consistent picture where C/O is enhanced.
More locally, the C/O ratio in a protoplanetary envelope could be enhanced by selective removal of water ice during pebble accretion \citep{2023MNRAS.523.6186W}, which is directly related to the mechanism discussed in Section \ref{sec:pa}.
Since the C/O ratio is a key parameter linking planet formation and exoplanet compositions \citep{2011ApJ...743L..16O, 2017MNRAS.469.4102M}, future observations and chemical modeling of the SiS source are crucial.

\section{Summary} \label{sec:sum}

In this Letter, we presented observations of the $\rm ^{28}SiS$ and $\rm ^{30}SiS$ lines in the protoplanetary disk around PDS 66. Our main results are summarized as follows:

\begin{itemize}

\item The \isis\ and \jsis\ lines are detected at $r=60$ au in the southwestern region of the disk. These constitute the second and first detections of $\rm ^{28}SiS$ and $\rm ^{30}SiS$, respectively, in protoplanetary disks. Both lines are spatially and spectrally co-located within uncertainties. The source is unresolved, located at the edge of the millimeter-continuum disk and within the infrared gap, and its velocity is consistent with Keplerian rotation, suggesting a connection to planet formation processes.

\item Modeling of the line fluxes constrains the column densities and emitting radius as a function of temperature. The detection of $\rm ^{30}SiS$ implies a SiS mass of $10^{22}{-}10^{23}\ {\rm g}$. The emitting radius derived from the optically thick main isotopologue emission is $0.5$–$4$ au.

\item The large fraction of silicon in SiS relative to the silicon inferred from dust grains at the same location indicates that, if the silicon originates from in situ processing of dust, complete sublimation of dust grain cores would be required. Alternatively, if silicon is locally accumulated relative to dust grains, sublimation of grain mantles alone would suffice.

\item We propose a circumplanetary envelope around a low-mass (${<}10\ M_\oplus$) protoplanet as a plausible explanation. Pebble accretion may supply silicon to the envelope through sublimation of grain cores and/or mantles. The high SiS/SiO ratio implied by the non-detection of SiO would be further related to an elevated C/O ratio in the planet-forming material if the chemistry in the envelope is similar to that of carbon stars.

\end{itemize}

The next step is to characterize the physical properties and molecular abundances of the source with higher spatial and spectral resolution observations.
A dedicated line survey and the ALMA wide-band sensitivity upgrade \citep{2023pcsf.conf..304C} will be pivotal in this endeavour.

\begin{acknowledgments}
We thank the anonymous referee for helpful comments and suggestions.
We would like to thank Kazumasa Ohno and Satoshi Okuzumi for fruitful discussions on pebble accretion and Tomoya Hirota for helpful suggestions on masers.
This Letter makes use of the following ALMA data: ADS/JAO.ALMA \# 2023.1.00334.S, 2023.1.00525.S, and 2017.1.01687.S.
ALMA is a partnership of ESO (representing its member states), NSF (USA), and NINS (Japan), together with NRC (Canada), NSTC and ASIAA (Taiwan), and KASI (Republic of Korea), in cooperation with the Republic of Chile.
The Joint ALMA Observatory is operated by ESO, AUI/NRAO, and NAOJ.
T.C.Y. was supported by the ALMA Japan Research Grant of NAOJ ALMA Project, NAOJ-ALMA-378.
This work was supported by Grant-in-Aid for JSPS Fellows, JP23KJ1008 (T.C.Y.).
F.A. and S.F. are funded by the European Union (ERC, UNVEIL, 101076613). Views and opinions expressed are however those of the authors only and do not necessarily reflect those of the European Union or the European Research Council. Neither the European Union nor the granting authority can be held responsible for them. S.F. also acknowledges financial contribution from PRIN-MUR 2022YP5ACE.
G.R. acknowledges support from the European Union (ERC Starting Grant DiscEvol, project number 101039651) and from Fondazione Cariplo, grant No. 2022-1217. Views and opinions expressed are, however, those of the author(s) only and do not necessarily reflect those of the European Union or the European Research Council. Neither the European Union nor the granting authority can be held responsible for them.
Support for C.J.L. was provided by NASA through the NASA Hubble Fellowship grant No. HST-HF2-51535.001-A awarded by the Space Telescope Science Institute, which is operated by the Association of Universities for Research in Astronomy, Inc., for NASA, under contract NAS5-26555.
MB has received funding from the European Research Council (ERC) under the European Union’s Horizon 2020 research and innovation programme (PROTOPLANETS, grant agreement No. 101002188).
L.P.  acknowledges support by the ANID BASAL project FB210003 and ANID FONDECYT Regular 1221442. 
\end{acknowledgments}

\facilities{ALMA}

\software{astropy\citep{astropy:2013, astropy:2018, astropy:2022},  CASA\citep{2022PASP..134k4501C}}

\appendix

\section{Observational Details} \label{app:obs}

We describe the observations and data reduction procedures here.
The data were processed using the Common Astronomy Software Application Modular version 6.5 \citep[CASA;][]{2022PASP..134k4501C} as described below.

\subsection{\isis}
The \isis\ line is located near the $\rm H_2CO\ 4_{0,4}-3_{0,3}$ line and was observed as part of \citet{2025ApJ...984L..18T} (project ID: \# 2023.1.00334.S, PI: L. Trapman). We used the data already reduced by them and summarized the main observational parameters here.
Observations were performed between 2023 October and 2024 May in ALMA Band 7, with baseline lengths of 15–984 m. The on-source integration time was 34 min, and the velocity resolution was $1.2\ \kms$. Pipeline calibration and self-calibration were performed following \citet{2025ApJ...984L...7L}. The continuum was subtracted in the visibility plane.
We produced image cubes using the CLEAN algorithm with a pixel size of $0\farcs02$ and  two channel widths, $\Delta v = 1.25\ \kms$ and $5.0\ \kms$, hereafter referred to as narrow- and wide-channel cubes, respectively.
The motivation for creating the wide-channel cube is to properly perform CLEAN by improving the S/N per channel for a more accurate flux estimate.
Natural weighting, the Hogbom deconvolver, and a CLEAN threshold of $3\sigma$ (RMS measured in line-free channels) were used. 
We did not apply a specific CLEAN mask but allowed the algorithm to construct CLEAN components for any signal above the threshold not to bias the algorithm.
The resultant beam size and RMS for the narrow-channel cube were ${\sim}0\farcs5$ and 1.1 $\rm mJy\ beam^{-1}$, respectively (Table \ref{tab:sis_obs}). The RMS for the wide-channel cube was 0.85 $\rm mJy\ beam^{-1}$.

\subsection{\jsis}
The \jsis\ line was observed in ALMA Cycle 11 under project ID \# 2023.1.00525.S (PI: T. Yoshida) on 2024 May, close in time to the second \isis\ execution. Baselines ranged 15–741 m, with an on-source integration time of 48 min. The line was included in a spectral window primarily targeting the continuum, with a velocity resolution of ${\sim}0.9\ \kms$.
Pipeline calibration included automatic self-calibration using the continuum image. Seven rounds of phase-only self-calibration were performed, improving the peak signal-to-noise ratio from 330 to 1700, and producing a nearly Gaussian noise distribution. The continuum was subtracted in the visibility plane, and visibility cubes were generated.
Image cubes were created using the same CLEAN procedure as for \isis, but only with a velocity channel width of $\Delta v = 1.25\ \kms$. The resulting beam size and RMS are ${\sim}0\farcs5$ and 1.7 $\rm mJy\ beam^{-1}$ (Table \ref{tab:sis_obs}).

\subsection{Channel Maps} \label{app:chanmap}
\begin{figure*}
  \centering
  \includegraphics[width=\textwidth]{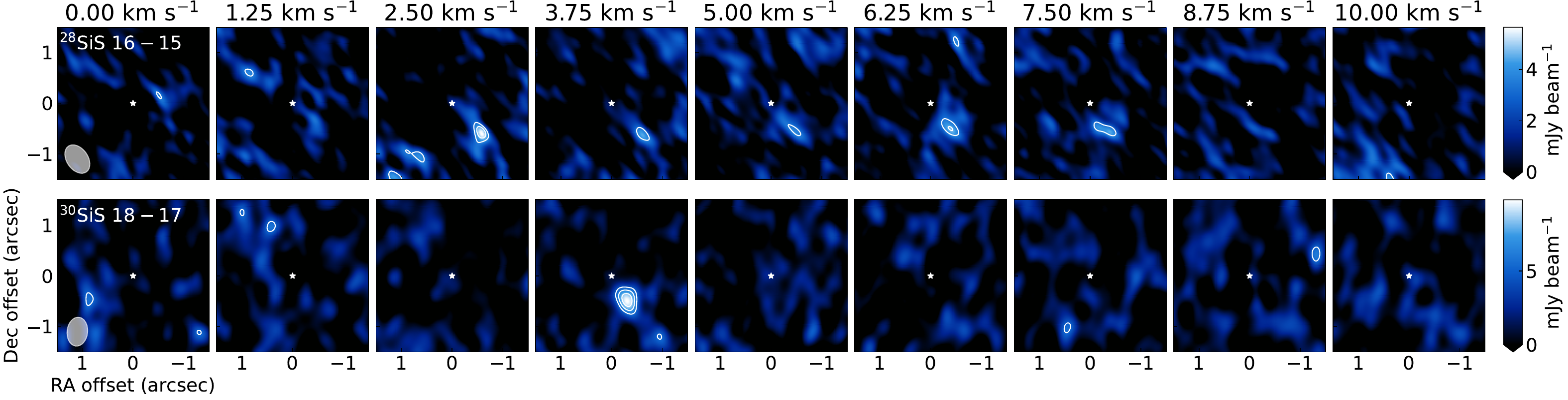}
  \caption{Channel maps of the \isis\ and \jsis\ lines are shown. The cubes with narrow channel widths are plotted. The contours are drawn at $[3,4,5]\sigma$ levels.} Star symbols indicate the position of the central star. The grey ellipses show the synthesized beams.
  \label{fig:chanmap}
\end{figure*}
Channel maps of the narrow-channel cubes for both \isis\ and \jsis\ lines are shown in Figure \ref{fig:chanmap}. The star symbols indicate the position of the central star.

\subsection{Histograms of the S/N Maps} \label{app:hist}

\begin{figure}
\centering
\includegraphics[width=0.4\textwidth]{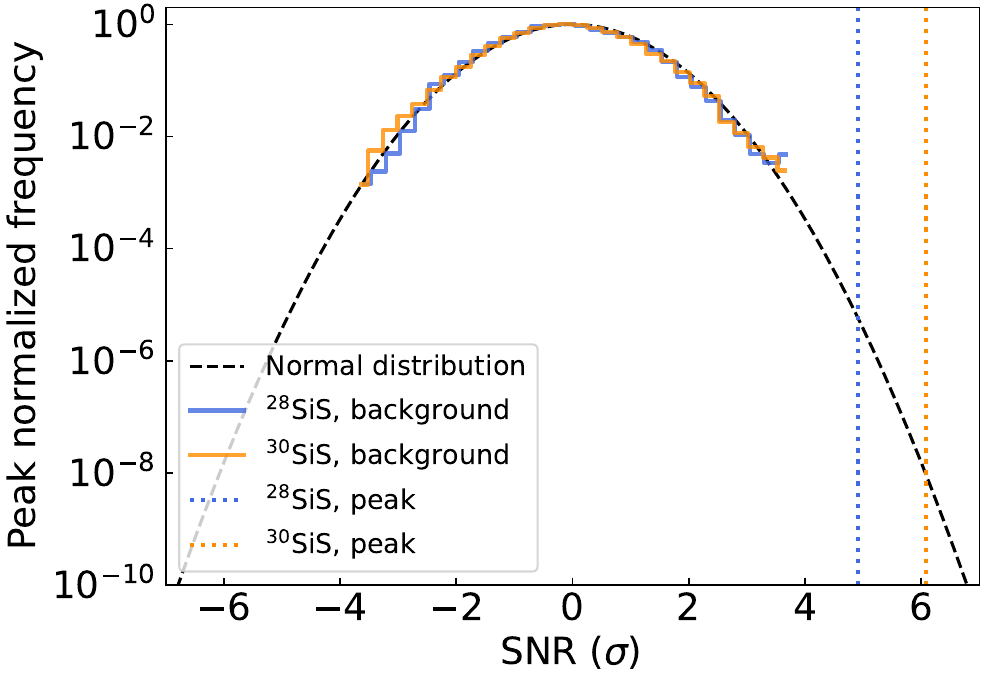}
\caption{Histograms of the background pixel values of the S/N maps. Solid colored lines show the normalized histograms, while the dotted colored lines indicate the peak values corresponding to the detected signals. The black dashed line shows a normal distribution normalized to its peak, which matches the histograms well.}
\label{fig:hist}
\end{figure}

Figure \ref{fig:hist} presents the histograms of the background pixel values in the S/N maps (Figure \ref{fig:mom0}).
All pixels within a $10''$ square centered on the star were included, excluding the central $1''$ radius region that contains the signal.
The histograms closely follow a Gaussian distribution and has no heavy tail which is sometimes seen in an image with insufficient calibration.
The detected signals are significantly above the background distribution.

\section{Caveats on Line Flux Modeling} \label{app:caveats}

To derive the physical quantities of SiS (Section \ref{sec:res2}), we assumed that the emission originates from a homogeneous slab and used the integrated fluxes.
However, this assumption might be inadequate if the source has complex density, temperature, or velocity structures.
Indeed, the line widths of the two isotopologues may differ (Figure \ref{fig:spec}); the \isis\ line has a width of ${\sim}5\ \kms$, while the \jsis\ line is unresolved at a velocity resolution of ${\sim}1\ \kms$, although the current signal-to-ratio of the \isis\ narrow-channel cube is not enough to conclude this.
Furthermore, the peak intensity of the \jsis\ line is about twice that of \isis.
This inversion may indicate a complex physical structure.
For example, cold, optically thick $\rm ^{28}SiS$ could obscure a hot inner component, while $\rm ^{30}SiS$, with lower optical depth, would allow the inner hot emission to emerge. 
We note that such an inverted temperature structure has been predicted for protoplanets \citep[e.g.,][]{2018A&A...611A..65B}.

Another caveat is the assumption of local thermal equilibrium (LTE).
The critical density of the \isis\ line is ${\sim}7 \times 10^{7}\ {\rm cm^{-3}}$ according to the collisional rates in the LAMDA database \citep{2005A&A...432..369S, 2008MNRAS.390..239K}.
We estimated the midplane gas density at $r=60$ au based on the gas mass derived from CO and $\rm N_2H^+$ emission \citep{2025ApJ...989....5T, 2025ApJ...984L..18T} and the characteristic radius from the rotation curve analysis \citep{2025ApJ...984L..17L}, using the parametric tapered power-law profile.
The gas surface density at $r=60$ au is estimated to be $\Sigma_g \sim 5\ {\rm g\ cm^{-2}}$.
The gas scale height is given by
\begin{equation} \label{eq:Hg}
H_g = \sqrt{ \frac{k_B T_{\rm mid} r^3 }{ \mu_{\rm gas} m_p G M_\star } } \simeq 3\ {\rm au},
\end{equation}
where $k_B$, $T_{\rm mid}$, $\mu_{\rm gas}$, $m_p$, and $G$ are the Boltzmann constant, midplane temperature, mean molecular weight of the gas, proton mass, and gravitational constant.
We adopt $\mu_{\rm gas} = 2.37$ and
\begin{equation} \label{eq:Tmid}
T_{\rm mid} = \left( \frac{0.02 L_\star}{ 8 \pi r^2 \sigma_{\rm SB} } \right)^{1/4} \simeq 17\ {\rm K},
\end{equation}
where $L_\star = 1.2\ L_\odot$ \citep{2023A&A...673A..77R} and $\sigma_{\rm SB}$ is the Stefan-Boltzmann constant, assuming stellar irradiation sets the midplane temperature \citep[e.g.,][]{2018ApJ...869L..42H}.
The resulting midplane gas density is $n_{\rm H_2} \sim 1\times10^{10}\ {\rm cm^{-3}}$, two orders of magnitude higher than the critical density.
Thus, the lines should be thermalized if collisional and spontaneous processes dominate the excitation.

Still, a possibility remains that the \jsis\ emission is masing.
The narrow line width and high peak flux could imply a maser, though they are also consistent with thermal broadening.
$^{28}$SiS masers are observed under similarly dense conditions around the carbon star IRC+10216 \citep{2006ApJ...646L.127F, 2017ApJ...843...54G}, where infrared pumping due to overlapped lines between the $^{28}$SiS ro-vibrational lines and other moleculer lines such as $\rm C_2H_2$ is a possible mechanism.
This could be relevant for PDS 66 if a strong infrared source (e.g., a protoplanet) is present.
However, the lack of collisional rate data and high-excitation spectroscopic data for \jsis\ prevents detailed modeling.

Future high-resolution observations and detailed modeling would be valuable.
Nevertheless, all the above scenarios are consistent, to varying degrees, with the presence of a local heating source and the sublimation of dust grains.

\section{Comments on the Catalogue Frequencies} \label{app:freq}

\begin{figure}
\centering
\includegraphics[width=0.4\textwidth]{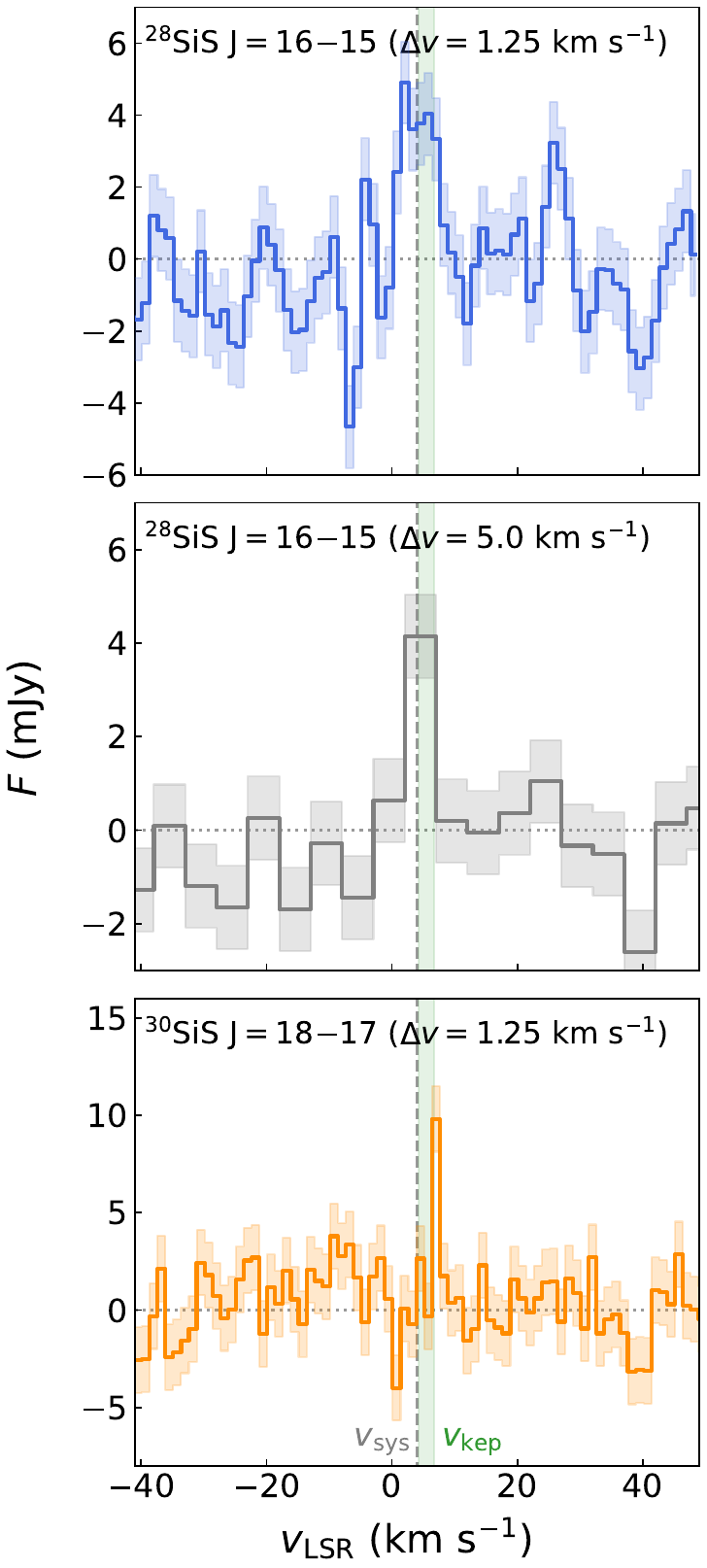}
\caption{Spectra of the SiS lines assuming the JPL database frequencies.}
\label{fig:spec_JPL}
\end{figure}

We found that the \isis\ and \jsis\ lines exhibit slightly different line-of-sight velocities.
One possibility is that the gas velocities are intrinsically different, which could arise from complex physical structures along the line of sight (Appendix \ref{app:caveats}).

Another possible explanation is an uncertainty in the spectroscopic data. In Figure \ref{fig:spec_JPL}, we show the spectra assuming the frequencies from the JPL database \citep{1998JQSRT..60..883P} instead of the CDMS database \citep{2007PCCP....9.1579M}.
While the \isis\ line remains unchanged, the \jsis\ line is redshifted by ${\sim}3\ \kms$ compared to the CDMS-based spectrum (Figure \ref{fig:spec}).
Because this transition was not directly measured experimentally but predicted from other transitions in the catalogues, the spectroscopic parameters might carry some uncertainty. To test this, we refer to a spectral survey of the carbon star IRC+10216 by \citet{2011ApJS..193...17P}, in which seven $\rm ^{30}SiS$ transitions, including \jsis, between 294–350 GHz were detected with the Submillimeter Array. They measured the line frequencies and compared them to the CDMS database. Among these lines, \jsis\ shows the largest blueshift, ${\sim}1$ MHz (or ${\sim}1\ \kms$), from the mean value, although the scatter is considerable.
This supports the possibility that the CDMS frequency for \jsis\ might be slightly overestimated. Future laboratory measurements are required to confirm this.



\bibliography{sample701}{}
\bibliographystyle{aasjournalv7}



\end{document}